\documentclass[trackchanges, twocolumn]{aastex701}
\usepackage{amsmath, amssymb, bm, multirow}

\begin{document}

\title{Modeling Dependence Structures in Astronomical Multi-Band Time Series Data\\via Multi-Output Gaussian Processes}

\correspondingauthor{Hyungsuk Tak}\email{tak@psu.edu}

\author[0009-0009-5730-9424]{Samata Das}
\affiliation{Department of Statistics,  Pennsylvania State University, University Park, PA 16802, USA}
\email{smd7237@psu.edu}
\author{Lishan Shi}
\affiliation{Department of Statistics,  Pennsylvania State University, University Park, PA 16802, USA}
\email{lfs5712@psu.edu}
\author[0000-0002-0957-7151]{Yasaman Hamayouni}
\affiliation{Department of Astronomy and Astrophysics,  Pennsylvania State University, University Park, PA 16802, USA}
\email{ybh5251@psu.edu}
\author[0000-0003-0334-8742]{Hyungsuk Tak}
\affiliation{Department of Statistics,  Pennsylvania State University, University Park, PA 16802, USA}
\affiliation{Department of Astronomy and Astrophysics,  Pennsylvania State University, University Park, PA 16802, USA}
\affiliation{Institute for Computational and Data Sciences,  Pennsylvania State University, University Park, PA 16802, USA}
\affiliation{The Research Institute of Basic Sciences, Seoul National University, Seoul 08826, Republic of Korea}
\email{tak@psu.edu}
\author{
Jong-Hak Woo}
\affiliation{Department of Physics and Astronomy, Seoul National University, Seoul 08826, Republic of Korea}
\email{woo@astro.snu.ac.kr}



\begin{abstract}
Modern astronomical time-domain surveys routinely collect multi-band light curves that provide complementary information about the physical processes governing source variability. \textcolor{black}{Although various multi-output Gaussian process (GP) models have been applied in different astronomical contexts, the statistical and astrophysical implications of their dependence structures have received comparatively limited attention. In this work, we examine three existing approaches to modeling cross-band dependence within a common multi-output GP framework and investigate how their underlying assumptions affect statistical inference and astrophysical interpretation. These include covariance-based, stochastic differential equation (SDE)-based, and latent-process approaches, which differ mainly in how cross-band dependence is introduced. The covariance-based approach directly specifies cross-band covariance functions to characterize dependence, whereas the SDE-based approach specifies joint temporal dynamics, with cross-band dependence arising from coupled dynamics or correlated driving processes. The latent-process approach represents dependence through shared stochastic processes and their transformations, which can reflect physical relationships among multiple light curves.} We demonstrate the practical implications of these modeling choices through applications to multi-band active galactic nucleus light curves and emission-line reverberation mapping. Rather than advocating a universally preferred approach, this work provides guidance for selecting dependence structures according to scientific objectives and clarifies how these choices influence the statistical characterization and scientific interpretation of multi-band stochastic variability.

\end{abstract}

\keywords{\uat{Quasars}{1319} --- \uat{Astrostatistcs}{1882} --- \uat{Astronomy Data Analysis}{1858}}



\section{Introduction}
\label{sec:intro}

Modern astronomical time-domain surveys have transformed the study
of variable and transient sources by repeatedly observing the sky
across multiple  bands. Large-scale surveys such as the
Sloan Digital Sky Survey (SDSS; \citealt{york2000}), the Zwicky
Transient Facility (ZTF; \citealt{bellm2019}), and the Vera C.
Rubin Observatory Legacy Survey of Space and Time (LSST;
\citealt{ivezic2019}) routinely produce multi-band light curves for
millions of astronomical objects. Since observations at different wavelengths provide
complementary information about the underlying physical processes, jointly modeling multi-band light curves has become increasingly important for studying variability in active galactic nuclei (AGNs) and variable stars, as well as transient phenomena such as supernovae
and tidal disruption events. However, individual bands are often modeled independently, thereby ignoring their dependence.

\textcolor{black}{Multi-output} Gaussian processes  \citep[GPs,][]{goovaerts1997,gelfand2010,cressie2011,alvarez2012} \textcolor{black}{provide a flexible probabilistic framework for jointly modeling multiple light curves while accommodating irregular sampling, heteroscedastic measurement uncertainties, and missing observations in some bands. Astronomical applications include modeling the stochastic variability of active galactic nuclei (AGNs) using multivariate damped random walk (DRW) and continuous-time autoregressive moving-average (CARMA) processes \citep{2020AJ....160..265H,schmidlkofer2026mcarma,schmidlkofer2026mcarma2},} 
reverberation mapping \textcolor{black}{between continuum and emission-line light curves} \citep{zu2016app,2025ApJ...992..130Y,li2024},
exoplanet detection \textcolor{black}{through joint modeling of radial velocities and stellar activity indicators}
\citep{haywood2014,rajpaul2015,jones2022,2023MNRAS.522.3458B},
and transient classification \textcolor{black}{via simultaneous modeling all photometric bands}
\citep{boone2019}.


\textcolor{black}{
These applications employ different approaches to representing
dependence among multiple light curves, reflecting their distinct
scientific objectives. Understanding the relationships among these
approaches, their statistical properties, and their implications
for astronomical inference is important for selecting appropriate
dependence structures for different astronomical applications.
In this work, we bring together three existing approaches to dependence
modeling within a unified multi-output GP framework, allowing us
to examine how different representations of dependence affect
the statistical characterization and scientific interpretation
of multi-band variability. Specifically, we consider covariance-based,
SDE-based, and latent-process approaches. Our goal is not to advocate
a universally preferred approach, but to clarify the modeling
choices and their consequences for scientific interpretation.}

\textcolor{black}{Although the three approaches are not mutually exclusive and may admit equivalent representations, they differ primarily in how cross-band dependence is introduced. The covariance-based approach specifies dependence directly through cross-covariance functions, subject to constraints ensuring a valid joint covariance structure. The SDE-based approach introduces dependence through coupled stochastic dynamics or correlated driving processes, while the latent-process approach introduces dependence through shared latent GPs and their transformations into individual light curves. These different constructions provide complementary ways to characterize cross-band dependence, with implications for the statistical properties and physical interpretation of astronomical multi-band variability.}

\begin{figure}
\begin{center}
\includegraphics[scale = 0.3]{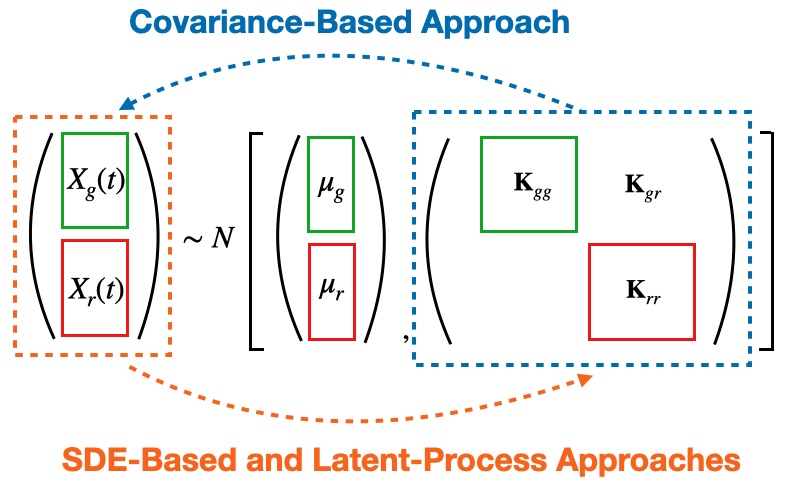}
\caption{
Conceptual illustration of the direction of model construction
in three approaches to dependence modeling in multi-output GPs.
In the covariance-based approach (top), the specified covariance
matrix determines the joint stochastic behavior of multiple
light curves. In the SDE-based and latent-process approaches
(bottom), the modeled stochastic processes induce the
corresponding covariance matrix. The dashed arrows indicate
these two directions of model construction.
}
\label{fig0}
\end{center}
\end{figure}

Figure~\ref{fig0} illustrates the two directions of model
construction underlying the three approaches. The GP values
in the $g$ and $r$ bands, evaluated at their respective
observation times, form a multivariate Gaussian random vector
with a corresponding mean vector and covariance matrix.
The diagonal blocks of the covariance matrix describe
temporal dependence within individual bands, whereas the
off-diagonal blocks characterize cross-band dependence.
In the covariance-based approach (top), specifying the
covariance matrix determines the joint stochastic behavior
of the multiple light curves. In contrast, the SDE-based
and latent-process approaches (bottom) begin by specifying
stochastic dynamics or latent-process representations,
respectively, which then induce the covariance matrix.
The dashed arrows highlight this distinction in the
direction of model construction.

\textcolor{black}{
We illustrate the statistical and astronomical implications
of dependence-structure modeling through two representative
applications: multi-band AGN variability and continuum
reverberation mapping. In the first application, we compare
a covariance-based multivariate Mat\'ern model with an
SDE-based multivariate damped random walk (mDRW) model
fitted to five-band AGN light curves. Despite yielding
similar marginal power spectral densities, the two models
imply substantially different cross-band coherence
structures. In the second application, we examine
latent-process models for continuum reverberation mapping
using top-hat and Gaussian transfer functions. 
These models yield different emission-line power spectral
densities and continuum--line cross-covariance structures,
while both imply unit squared coherence between the
latent continuum and emission-line processes, meaning
that the emission-line variability is entirely driven
by the latent continuum through a deterministic linear
response. Together,
these applications demonstrate how dependence-structure
assumptions affect statistical inference and the
interpretation of astronomical variability.
}

The remainder of this paper is organized as follows.
Section~\ref{sec2} introduces the covariance-based,
SDE-based, and latent-process approaches within a unified
multi-output GP framework, together with a likelihood
formulation for irregularly sampled observations with
heteroscedastic measurement uncertainties.
Section~\ref{sec3} develops their time- and frequency-domain
representations, using multivariate DRW models as
illustrative examples, and discusses the resulting
statistical properties.
Section~\ref{sec4} investigates the astronomical implications
of dependence-structure modeling through applications
to multi-band AGN variability and continuum reverberation
mapping. Finally, Section~\ref{sec5} summarizes the
main findings and discusses  future directions.

\section{Dependence Structures in Multi-Output Gaussian Processes}
\label{sec2}

We begin by introducing a common statistical framework for multi-output GPs based on the multivariate Gaussian distribution of the observed light curves. Although the likelihood function takes the same general form, the covariance matrix depends on how within- and cross-band dependence is specified. We then describe three approaches to modeling this dependence: covariance-based, stochastic differential equation (SDE)-based, and latent-process approaches.

Throughout this manuscript, let
$$
\mathbf X(t)
=
\left(
X_1(t),\ldots,X_k(t)
\right)^\top
$$
denote a $k$-dimensional multi-output GP describing the
underlying stochastic variability of the $k$ light curves
at time $t$.
We assume
\[
\mathbf X(\cdot)
\sim
\mathcal{GP}
\left(
\boldsymbol{\mu}(\cdot),
\mathbf K(\cdot,~\cdot)
\right),
\]
where $\boldsymbol{\mu}(t)
=
E\{\mathbf X(t)\}
$ and $
\mathbf K(t,t')
=
{\rm Cov}
\left(
\mathbf X(t),
\mathbf X(t')
\right)
$
is a valid matrix-valued covariance function.

Suppose observations across all bands are obtained at \(n\) distinct time points \(t_1,\ldots,t_n\), where one or more  bands may be observed at each observation time. Let $x_{ij}$ denote the observed
brightness in  band $j$ at time $t_i$.  We assume
that the corresponding measurement uncertainty, $\delta_{ij}$,
is known for each observation, as is typically provided by modern
astronomical time-domain surveys.

We assume the Gaussian measurement error model
\[
x_{ij}
=
X_j(t_i)
+
\varepsilon_{ij},
\]
where $X_j(t_i)$ denotes the latent  GP value in  band $j$ evaluated at time $t_i$,
and the measurement errors
\[
\varepsilon_{ij}
\sim
N(0,~  \delta_{ij}^2)
\]
are independent both across observations and from the latent GP.
Let \(\mathbf x\) denote the vector obtained by stacking all available observations, and let \(\mathbf X\) denote the corresponding vector of latent GP values, obtained by stacking the components of \(\mathbf X(t_1),\ldots,\mathbf X(t_n)\) for which measurements are available, in the same order as \(\mathbf x\). We define
$
\boldsymbol{\mu}=E(\mathbf X)$ and $\mathbf K=\operatorname{Cov}(\mathbf X).
$
Furthermore, let
\[
N
=
\dim(\mathbf x)
=
n_1+\cdots+n_k
\]
denote the total number of observations, where $n_i$ is the
number of observations in  band $i$.  Under the
Gaussian measurement error model, 
the observed data  follow a multivariate Gaussian
distribution,
\[
\mathbf x\sim\mathcal N(\boldsymbol\mu,~ \mathbf K+D_\delta).
\]
where $D_\delta
=
\mathrm{diag}
\left(
\delta_1^2, \ldots, \delta_N^2
\right)$ is the measurement-error covariance matrix with \(\delta_1,\ldots,\delta_N\) denoting the measurement uncertainties arranged in the same order as the observations in \(\mathbf x\).
Denoting $\Sigma=\mathbf{K}+D_\delta$, the  log-likelihood is given by
$$
\ell =
-\frac12\left[
N\log(2\pi)+\log|\Sigma|
+(\mathbf x-\boldsymbol\mu)^\top
\Sigma^{-1}(\mathbf x-\boldsymbol\mu)
\right].
$$

Maximum likelihood, Bayesian inference, and simulation-based inference can all be applied within this common probabilistic framework. The three approaches considered here differ primarily in how cross-band dependence is introduced and the resulting covariance matrix $\mathbf{K}$ is constructed, rather than in the inferential methodology itself. Our emphasis is therefore on the specification and scientific interpretation of dependence structures rather than computational strategies for GP inference. Computational issues, including scalable covariance approximations and efficient algorithms for large GP models, are beyond the scope of this work; interested readers are referred to \citet{rasmussen2006,foremanmackey2017,heaton2019,liu2020}.


\subsection{Covariance-Based Approach}
\label{sec21}

The covariance-based approach \citep{gelfand2010,cressie2011,alvarez2012} constructs a multi-output GP by
specifying its matrix-valued covariance function directly. More
specifically, dependence among  bands is represented
through cross-covariance functions, while the stochastic
variability within each band is characterized by its own
covariance kernel. Consequently, the principal mathematical
challenge is to construct matrix-valued covariance functions that
remain jointly positive definite over both  bands and
observation times.

Let
\begin{equation}\label{eq:covmat}
\mathbf K(t,t')
=
\left\{
K_{ij}(t,t')
\right\}_{i,j=1}^{k}
\end{equation}
denote the matrix-valued covariance function of the multi-output
GP, where
$$
K_{ij}(t,t')
=
\mathrm{Cov}
\left(
X_i(t),X_j(t')
\right)
$$
is the cross-covariance between bands $i$ and $j$ at times $t$ and $t'$, respectively. The diagonal elements, $K_{ii}(t,t')$, characterize the variability and temporal dependence within individual bands through their marginal covariance functions, such as a DRW, Mat\'ern, periodic, or
quasi-periodic model, whereas the off-diagonal elements, $K_{ij}(t,t')$ for $i\neq j$, characterize cross-band dependence.

Combining individually valid marginal covariance kernels and positive-definite cross-band covariance functions, however, does not necessarily ensure the positive definiteness of the resulting covariance function of the multi-output GP jointly across bands and observation times. Constructing valid cross-covariance functions in Eq.~\eqref{eq:covmat} is therefore more challenging than specifying marginal covariance kernels. \textcolor{black}{One established covariance-based construction is the multivariate Mat\'ern process \citep{gneiting2010matern,apanasovich2012matern}, which allows band-specific temporal covariance kernels while imposing appropriate constraints to ensure the validity of the joint covariance function.}

\textcolor{black}{Specifically, the covariance functions of a multivariate Mat\'ern process take the form
$$
K_{ij}(t,t')
=
\rho_{ij}s_i s_j
\frac{2^{1-\nu_{ij}}}{\Gamma(\nu_{ij})}
\left(\frac{|t-t'|}{\ell_{ij}}\right)^{\nu_{ij}}
\mathcal K_{\nu_{ij}}
\left(\frac{|t-t'|}{\ell_{ij}}\right).
$$
where  $s_j$ denotes the stationary standard deviation of $X_j(t)$, $\nu_{ij}$ controls the smoothness, $\ell_{ij}$ is a temporal scale parameter,  $\mathcal K_{\nu_{ij}}$ denotes the modified Bessel function of the second kind. Also, 
$$
\rho_{ij}=\operatorname{Corr}\{X_i(t),X_j(t)\}=\frac{K_{ij}(t,t)}{s_i s_j}
$$ 
denotes the zero-lag correlation between the latent processes in bands $i$ and $j$, which is independent of $t$ under stationarity. The marginal covariance function of band~$j$ follows by setting $i=j$. The cross-band smoothness, temporal scale, and correlation parameters cannot generally be specified independently. Valid multivariate Mat\'ern constructions impose relationships among these parameters to ensure that the resulting matrix-valued covariance function is positive semidefinite \citep{gneiting2010matern,apanasovich2012matern}.}

\textcolor{black}{As a representative example, we consider the multivariate Mat\'ern construction of \citet{apanasovich2012matern}, specialized to $\nu_{ij}=1/2$ (i.e., DRW) for all pairs of bands. Under this specialization, the cross-covariance functions reduce to exponential kernels,
\begin{equation}
\label{eq:matern_drw_cross}
K_{ij}(t,t')
=
\rho_{ij}s_i s_j
\exp\left(-\frac{|t-t'|}{\tau_{ij}}\right),
\end{equation}
where
$$
\tau_{ij}
=
\left(
\frac{\tau_i^{-2}+\tau_j^{-2}}{2}
\right)^{-1/2},
\quad
\rho_{ij}
=
v_{ij}
\sqrt{
\frac{2\tau_i\tau_j}{\tau_i^2+\tau_j^2}
}.
$$
Here, $\tau_j$ is the characteristic timescale of band $j$, $\mathbf V=\{v_{ij}\}$ is an auxiliary positive-semidefinite correlation matrix used to ensure the validity of the multivariate Mat\'ern covariance function.}

\subsection{SDE-Based Approach}
\label{sec:sde}

\textcolor{black}{An alternative approach to constructing cross-band dependence is to specify a system of stochastic differential equations (SDEs) driven by correlated stochastic processes. Unlike covariance-based approaches, which directly specify the marginal and cross-covariance functions, SDE-based models induce these covariance functions through the underlying stochastic dynamics. This construction allows different bands to exhibit distinct temporal dependence structures while sharing correlated sources of stochastic variability.}

\textcolor{black}{Specifically, consider a $k$-dimensional process $\mathbf X(t)=(X_1(t),\ldots,X_k(t))^\top$ satisfying the linear SDE 
$$
d\mathbf X(t)
=
\mathbf A\{\mathbf X(t)-\boldsymbol\mu\}\,dt
+
\mathbf B\,d\mathbf W(t),
$$
where $\boldsymbol\mu$ is the long-term mean vector, $\mathbf A$ governs the temporal dynamics, $\mathbf B$ determines the stochastic variability, and $\mathbf W(t)$ is a vector of independent standard Wiener processes (Brownian motions). Cross-band dependence can arise through both the drift matrix $\mathbf A$, which governs interactions among bands, and the diffusion matrix $\mathbf B$, which determines the covariance structure of the stochastic driving terms. Under appropriate stability conditions, the system admits a stationary solution with a valid matrix-valued covariance function \citep{gardiner2009}.}

\textcolor{black}{As a representative example, we consider the multivariate damped random walk (DRW) model of \citet{2020AJ....160..265H}, which restricts the drift matrix $\mathbf A$ to be diagonal, allowing band-specific characteristic timescales while introducing cross-band dependence exclusively through correlated Brownian driving motions. Specifically, the process for band $j$ satisfies
$$
dX_j(t)
=
-\frac{1}{\tau_j}\{X_j(t)-\mu_j\}\,dt
+
\sigma_j\,dW_j(t),
$$
where $\mu_j$ is the long-term mean, $\tau_j>0$ is the characteristic timescale, and $\sigma_j>0$ is the diffusion coefficient of band $j$. The Brownian motions $W_j(t)$ are correlated such that
$$
\operatorname{Cov}\{dW_i(t),dW_j(t)\}
=
r_{ij}\,dt,
$$
where $\mathbf Q=\{r_{ij}\}$ is a positive-semidefinite correlation matrix of the driving Brownian  motions.}

\textcolor{black}{For the stationary multivariate DRW process, the marginal standard deviation is $s_j=\sigma_j\sqrt{\tau_j/2}$, and the resulting cross-covariance function is
\begin{equation}
\label{eq:mdrw_cross}
K_{ij}(t,t')
=
\rho_{ij}s_i s_j
\begin{cases}
\exp\left(-|t-t'|/\tau_i\right), & t\geq t',\\
\exp\left(-|t-t'|/\tau_j\right), & t<t',
\end{cases}
\end{equation}
where
$$
\rho_{ij}
=\operatorname{Corr}\{X_i(t),X_j(t)\}=
r_{ij}
\frac{2\sqrt{\tau_i\tau_j}}{\tau_i+\tau_j}.
$$
is the resulting zero-lag cross-band correlation  of the latent processes in band $i$ and $j$.}

\textcolor{black}{Unlike the multivariate Mat\'ern construction with $\nu_{ij}=1/2$ in Eq.~\eqref{eq:matern_drw_cross}, whose cross-covariance functions are symmetric in the time lag, the multivariate DRW cross-covariance function generally exhibits different decay rates for positive and negative time lags when $\tau_i\neq\tau_j$, as indicated in Eq.~\eqref{eq:mdrw_cross}. This time-lag asymmetry does not violate the symmetry requirement for a valid covariance function, $K_{ij}(t,t')=K_{ji}(t',t)$. For astronomical multi-band light curves, such asymmetry allows different temporal dependence patterns depending on which band leads or follows the other, although it does not necessarily imply a physical time delay or causal relationship between bands.}

\subsection{Latent-Process Approach}
\label{sec22}

Unlike the covariance-based and SDE-based approaches, the
latent-process approach introduces dependence indirectly through
one or more latent GPs and deterministic linear operators.
Consequently, the covariance function of the observed
multi-output GP is induced rather than specified explicitly.
Since GPs are closed under deterministic linear
transformations, the resulting covariance function is
automatically positive definite. This construction has been
widely adopted in machine learning, spatial statistics, and
astronomical time-series analysis because it naturally combines
mathematical validity with physically interpretable descriptions
of the mechanisms generating the observed data
\citep{alvarez2012}.

Let
\[
\mathbf Z(t)
=
\left(
Z_1(t),\ldots,Z_r(t)
\right)^\top
\]
denote $r$ latent GPs, i.e.,
\[
\mathbf Z(\cdot)
\sim
\mathcal{GP}
\left(
\boldsymbol{\mu}_Z(\cdot),
\mathbf K_Z(\cdot,~\cdot)
\right),
\]
with
$\boldsymbol{\mu}_Z(t)=\mathrm E(\mathbf Z(t))$ and $\mathbf K_Z(t,t')
=
\mathrm{Cov}
(
\mathbf Z(t),
\mathbf Z(t'))$.
The  multi-output GP is obtained by applying linear
operators to these latent processes,
\[
X_j(t)
=
\sum_{\ell=1}^{r}
\mathcal L_{j\ell}
\{
Z_\ell
\}(t),
\qquad
j=1,\ldots,k,
\]
where $\mathcal L_{j\ell}$ denotes a linear operator acting on the
$\ell$th latent GP ($\ell=1, \ldots, r$). Since GPs are
closed under linear transformations, the resulting multi-output
processes remain Gaussian.

This general formulation encompasses a broad class of
latent-process models because different choices of the linear
operators induce different dependence structures among the
observed  bands. For example, the simplest
construction assumes linear mixing,
$$
\mathcal L_{j\ell}\{Z_\ell\}(t)
=
a_{j\ell}Z_\ell(t).
$$
This formulation is
widely used in the machine learning and spatial statistics
literature on multi-output GPs and forms the basis of the linear
model of coregionalization and related latent-factor models
\citep{goovaerts1997,teh2005,alvarez2012}. In astronomy,
\citet{kelly2011} have employed an analogous linear-mixing
construction to represent a single-band light curve as a linear
combination of latent DRW processes, thereby obtaining a more
flexible PSD than that of a single DRW process.

Another important choice of linear operator arises in 
reverberation mapping, in which each observed light curve is
modeled as the convolution of one or more latent GPs with
wavelength-dependent transfer functions,
\begin{equation}\label{eq:rmlinoperator}
\mathcal L_{j\ell}\{Z_\ell\}(t)
=
\int
\Psi_{j\ell}(\tau)
Z_\ell(t-\tau)
\,d\tau,
\end{equation}
where $\Psi_{j\ell}(\tau)$ denotes the transfer function
\citep{li2024,cackett2021}. Because the convolution operator
satisfies the linearity property
\[
\mathcal L(aZ_1+bZ_2)
=
a\mathcal L(Z_1)
+
b\mathcal L(Z_2),
\]
it constitutes a linear operator acting on the latent GPs and
therefore fits naturally within the general latent-process
framework.

Similarly, GP models for stellar activity often
employ derivative operators,
\[
\mathcal L_{j\ell}\{Z_\ell\}(t)
=
a_{j\ell}Z_\ell(t)
+
b_{j\ell}
\frac{d}{dt}
Z_\ell(t),
\]
or related linear combinations of GPs and their
derivatives to jointly model radial velocity and stellar activity
indicators
\citep{rajpaul2015,gilbertson2020,jones2022, 2023MNRAS.522.3458B}.

These examples demonstrate that the distinction among latent-process models lies primarily in the choice of the linear operator rather than in the GP formulation itself. Consequently, just as the covariance kernel plays the central role in covariance-based models, the linear operator determines the covariance structure in latent-process models.

Since the linear operators are deterministic, the covariance function of the multi-output GP is obtained by applying the corresponding linear operators to the covariance functions of the latent processes. With mutually independent latent processes, the cross-covariance function is expressed as
$$
K_{ij}(t,t')
=
\sum_{\ell=1}^{r}
\mathcal L_{i\ell}
\mathcal L_{j\ell}
\operatorname{Cov}
\left(
Z_\ell(t),
Z_\ell(t')
\right),
$$
where the operators act on the first and second time arguments, respectively. Different choices of these operators induce different dependence structures among the observed light curves.

For example, in reverberation mapping, multiple observed light curves can be represented as convolutions of a common latent driving process with their respective transfer functions. Assuming a single latent process ($\ell=1$) in Eq.~\eqref{eq:rmlinoperator}, the cross-covariance function reduces to
\begin{equation}\label{eq:latent_crosscov}
K_{ij}(t,t')
=
\int\!\!\int
\Psi_i(u)\Psi_j(v)
K^{(Z)}(t-u,t'-v)
\,du\,dv,
\end{equation}
where $K^{(Z)}(t,t')=\operatorname{Cov}\{Z(t),Z(t')\}$ is the covariance function of the latent process. If the latent process follows a stationary DRW, its covariance function is
\[
K^{(Z)}(t,t')
=
\frac{\sigma^{2(Z)}\tau^{(Z)}}{2}
\exp\left(-\frac{|t-t'|}{\tau^{(Z)}}\right),
\]
where $\sigma^{(Z)}$ and $\tau^{(Z)}$ denote the diffusion coefficient and characteristic timescale of the latent DRW, respectively. Thus, cross-band dependence is determined jointly by the temporal covariance structure of the latent driving process and the transfer functions, which describe how the driving variability is reflected in the observed light curves.

The latent-process approach possesses several attractive
features for astronomical time-series analysis. First, dependence among  bands is interpreted through
shared latent stochastic processes rather than observable
cross-covariance functions, often leading to a more physically
interpretable description of the underlying variability
mechanisms.
Second, multiple observed light curves may be explained by a
relatively small number of latent GPs, leading to parsimonious
representations when the observed variability is believed to
arise from a limited number of common physical drivers. Finally,
because the covariance function is induced through deterministic
linear operators acting on latent GPs, the resulting
multi-output covariance function is automatically positive
definite, eliminating the need to construct valid
cross-covariance functions directly.

\section{Frequency-Domain Interpretation of Dependence Structures}
\label{sec3}

\textcolor{black}{The three approaches  specify different constructions of cross-band dependence within a common multi-output GP framework. For stationary multi-output GPs, the covariance functions and corresponding spectral densities are related through the Fourier transform. Specifically, the cross-spectral density between bands $i$ and $j$ is
$$
S_{ij}(\omega)
=
\int_{-\infty}^{\infty}
K_{ij}(h)e^{-i\omega h}\,dh,
$$
where $K_{ij}(h)=\operatorname{Cov}\{X_i(t+h),X_j(t)\}$ and $\omega$ denotes angular frequency. The marginal power spectral density (PSD) of band $i$,  $S_{ii}(\omega)$, is the special case when $i=j$. Thus, the within-band variability and cross-band dependence under all three approaches can be characterized within the same frequency-domain framework.}

\textcolor{black}{Under the covariance-based approach, consider the multivariate Mat\'ern construction with $\nu_{ij}=1/2$ in Eq.~\eqref{eq:matern_drw_cross}. Its exponential cross-covariance function yields
\begin{equation}\label{eq:psd_matern}
S_{ij}(\omega)
=
\frac{2\rho_{ij}s_i s_j\tau_{ij}}
{1+\omega^2\tau_{ij}^2}.
\end{equation}
For $i=j$, this reduces to the familiar Lorentzian PSD of a univariate DRW, with characteristic break frequency $\omega_b=\tau_i^{-1}$. The cross-spectra are real-valued because the cross-covariance functions are symmetric in the time lag. Consequently, this construction characterizes cross-band dependence across frequencies without introducing phase differences between bands.}

\textcolor{black}{Under the SDE-based approach, the multivariate DRW has cross-covariance functions with different decay rates for positive and negative time lags, as distinguished in Eq.~\eqref{eq:mdrw_cross}. Their Fourier transforms give
$$
S_{ij}(\omega)
=
\frac{\rho_{ij}s_i s_j(\tau_i+\tau_j)}
{(1+i\omega\tau_i)(1-i\omega\tau_j)}.
$$
Unlike the multivariate Mat\'ern construction in Eq.~\eqref{eq:psd_matern}, these cross-spectra are generally complex-valued when $\tau_i\neq\tau_j$. Their magnitudes characterize the strength of cross-band dependence at different frequencies, while their phases characterize frequency-dependent temporal asymmetry. The marginal PSDs remain Lorentzian, with band-specific characteristic timescales. Thus, the SDE-based construction allows frequency-dependent phase relationships even though the individual bands follow DRW processes.}

Under the latent-process approach, consider the reverberation-mapping model with a single latent driving process. The cross-spectral density is
\begin{equation}\label{eq:psd_latent}
S_{ij}(\omega)
=
\widehat{\Psi}_i(\omega)
\overline{\widehat{\Psi}_j(\omega)}
S_Z(\omega),
\end{equation}
where $\widehat{\Psi}_j(\omega)$ denotes the Fourier transform of the transfer function $\Psi_j$ and $S_Z(\omega)$ is the PSD of the latent driving process. If the latent process follows a stationary DRW,
\begin{equation}\label{eq:psd_latentdrw}
S_Z(\omega)
=
\frac{\sigma_Z^2\tau_Z^2}
{1+\omega^2\tau_Z^2}.
\end{equation}
The marginal PSDs are obtained by setting $i=j$, yielding $S_{ii}(\omega)=|\widehat{\Psi}_i(\omega)|^2S_Z(\omega)$. The transfer functions therefore determine how the driving variability is filtered across frequencies, while their relative phases characterize temporal relationships between the latent band-specific light curves $X_j(t)$. In reverberation mapping, these phase relationships can be interpreted in terms of the temporal response of the emitting regions to continuum variability.

These examples illustrate how the three approaches introduce cross-band dependence through different mechanisms, while their resulting dependence structures can all be characterized within the same frequency-domain framework. The covariance-based multivariate Mat\'ern construction specifies cross-spectral dependence directly through symmetric cross-covariance functions; the SDE-based multivariate DRW introduces frequency-dependent phase relationships through correlated stochastic driving processes and band-specific relaxation timescales; and the latent-process approach induces cross-spectra through shared driving variability and transfer functions. Their frequency-domain representations therefore provide complementary ways to investigate the statistical and astronomical implications of cross-band dependence. A detailed
derivation of the three cross-spectral densities is provided in Appendix~\ref{appendixa}.

\section{Illustrative Astronomical Applications}\label{sec4}

We investigate how different dependence structures affect the statistical characterization and astronomical interpretation of AGN variability through two illustrative applications. First, using multi-band AGN light curves, we compare the covariance-based multivariate Mat\'ern construction with $\nu_{ij}=1/2$ and the SDE-based multivariate DRW. Both models assume DRW marginal processes for each band, but differ in their cross-covariance functions and corresponding cross-spectral properties. This comparison allows us to examine the implications of different cross-band dependence structures while retaining the same marginal stochastic model. Second, using emission-line reverberation mapping, we illustrate the latent-process approach by considering different transfer functions that relate the driving continuum to the responding emission-line light curve. We examine how these choices affect the induced dependence structure and the interpretation of the continuum--line relationship.

\subsection{Multi-band AGN Variability}\label{sec41}

\begin{figure}
\begin{center}
\includegraphics[scale=0.18]{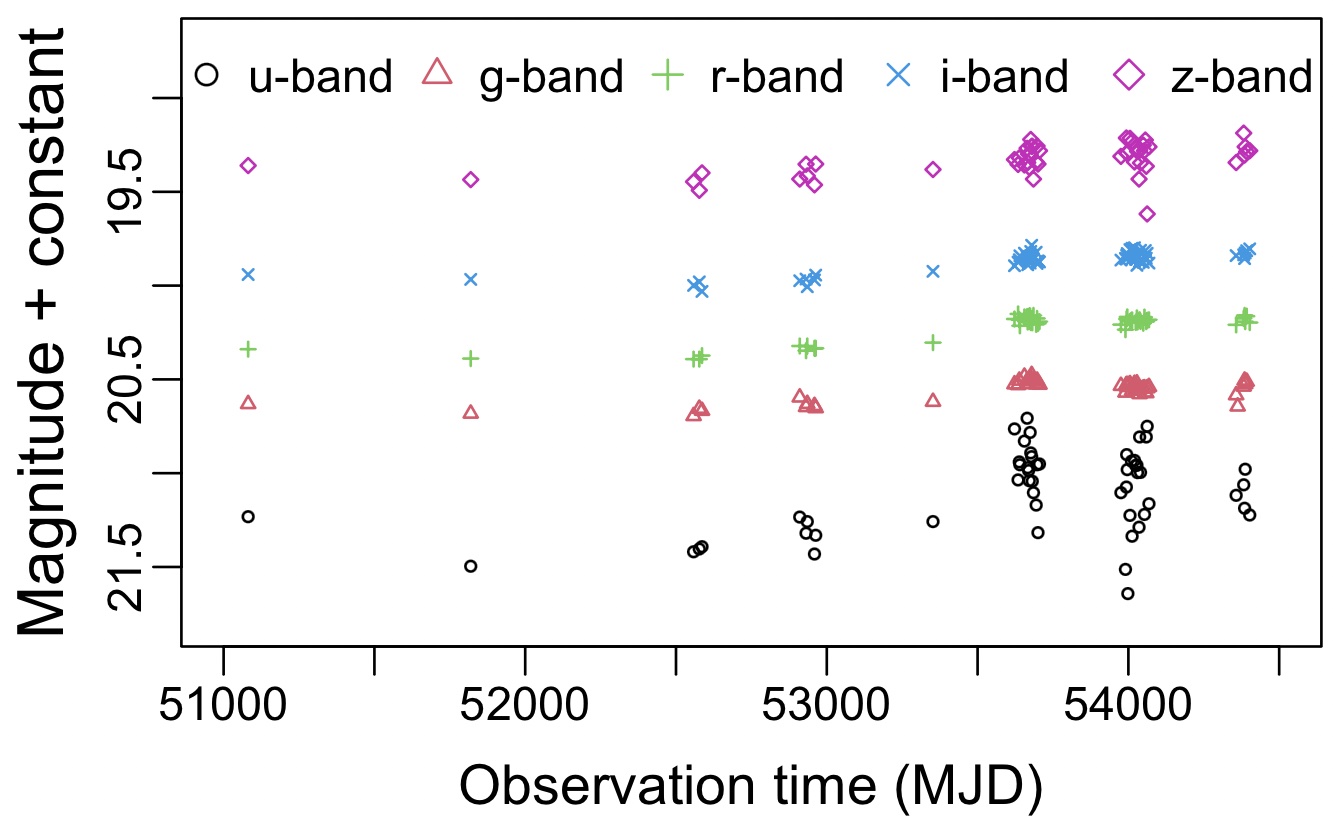}
\caption{
Five-band SDSS Stripe~82 light curves of a quasar (dbID: 1540),
selected from the catalog of 9258 spectroscopically confirmed
quasars compiled by \citet{macleod2012}. The observations are
irregularly sampled and contain different numbers of measurements
across the five photometric bands. The reported 1-$\sigma$
measurement uncertainties are smaller than the plotting symbols
and are therefore not visible.
}
\label{fig1}
\end{center}
\end{figure}

Stochastic variability in active galactic nuclei (AGNs) provides
a natural setting for comparing covariance-based and SDE-based
approaches to multi-output GP modeling, particularly when
investigating dependence among  light curves observed in
different photometric bands. We analyze one multi-band AGN
data set from the SDSS Stripe~82 survey \citep{york2000}.
This object (dbID: 1540) is taken from a catalog of 9258
spectroscopically confirmed SDSS Stripe~82 quasars
\citep{macleod2012}. The data consist of measurements in five
photometric bands, $u$, $g$, $r$, $i$, and $z$, together with
their reported measurement uncertainties. Table~\ref{tab:observation}
summarizes the number of observations, median observational
cadence, and median reported 1$\sigma$ measurement uncertainty
in each band. Figure~\ref{fig1} displays the corresponding
five-band light curves.

\begin{table}
\centering
\caption{
Observational summary of the five-band SDSS Stripe~82 quasar
(dbID: 1540). For each photometric band, the table lists the
number of observations, the median observational cadence (days),
and the median reported 1$\sigma$ measurement uncertainty.
}
\label{tab:observation}
{\begin{tabular}{lccccc}
\hline
 & $~~~u~~~$ & $~~~g~~~$ & $~~~r~~~$ & $~~~i~~~$ & $~~~z~~~$ \\
\hline
Number of obs. & 54 & 56 & 52 & 56 & 53 \\
Cadence (days) & 4.08 & 4.04 & 4.99 & 4.04 & 4.53 \\
1$\sigma$ error bar & 0.13 & 0.02 & 0.02 & 0.02 & 0.04 \\
\hline
\end{tabular}}
\end{table}

\begin{figure*}
\begin{center}
\includegraphics[width=\textwidth]{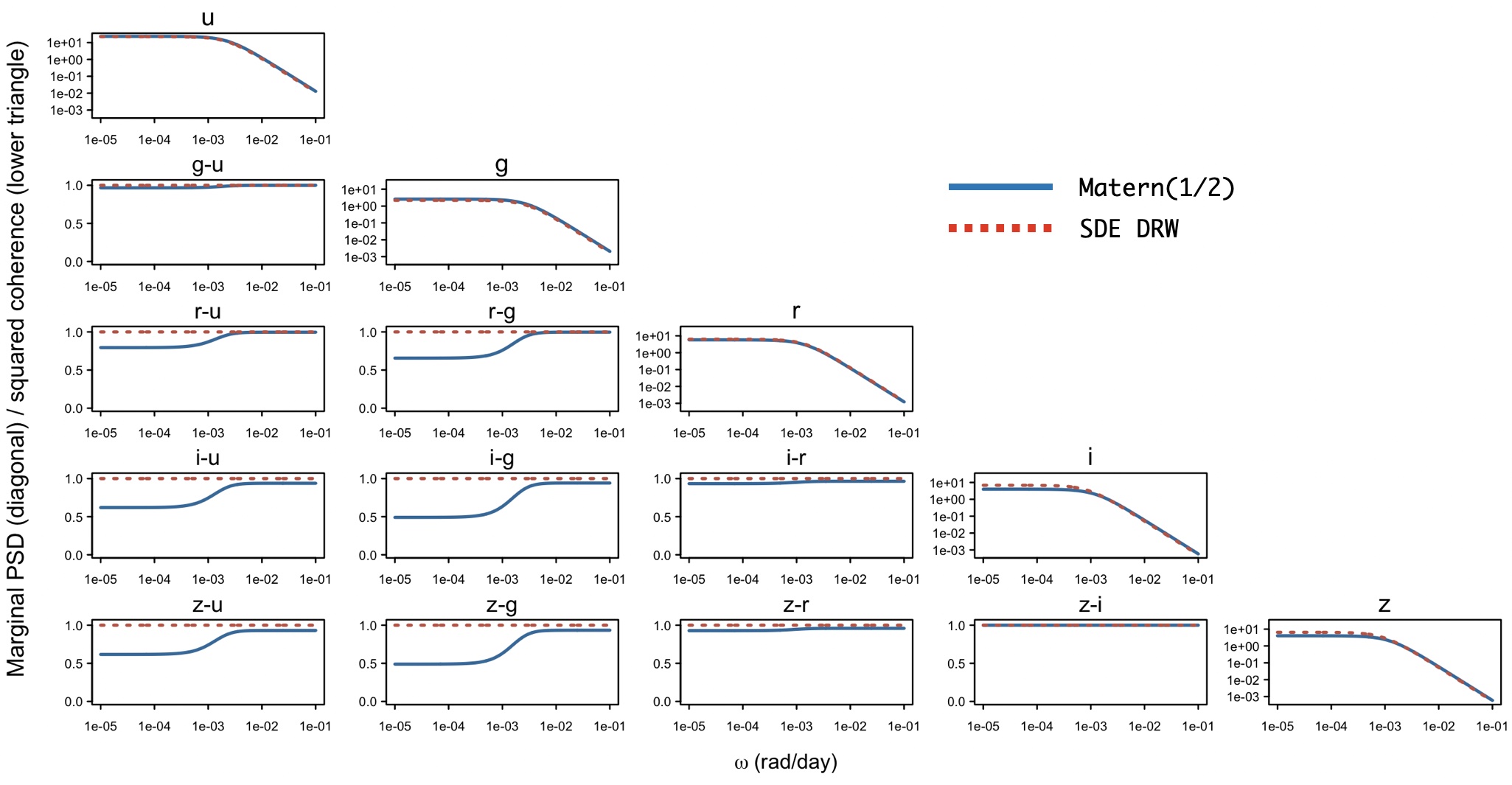}
\caption{
Comparison of the fitted covariance-based multivariate
Mat\'ern model ($\nu_{ij}=1/2$; solid blue) and SDE-based
multivariate DRW model (dashed red) for the SDSS Stripe~82
quasar (dbID: 1540). Diagonal panels display the marginal
power spectral densities, while lower-triangular panels
display squared coherence between pairs of photometric bands.
The marginal PSDs are nearly identical under the two models.
In contrast, the Mat\'ern model implies frequency-dependent
coherence, whereas the SDE-based mDRW model implies frequency-independent coherence close to unity.
}
\label{fig2}
\end{center}
\end{figure*}

\textcolor{black}{We fit the covariance-based multivariate Mat\'ern model with
$\nu_{ij}=1/2$ and the SDE-based multivariate DRW model described
in Sections~\ref{sec21} and~\ref{sec:sde}, respectively, using
maximum likelihood estimation. Both models assume DRW marginal
processes for each photometric band, but differ in their
cross-covariance structures, as shown in
Eqs.~\eqref{eq:matern_drw_cross} and~\eqref{eq:mdrw_cross}.
The parameter estimates and approximate Hessian-based  standard errors are
reported in Appendix~\ref{appendixb}. The estimated stationary
standard deviations and characteristic timescales are broadly
comparable between the two models. The most noticeable differences
occur in the $i$ and $z$ bands: the Mat\'ern model estimates
characteristic timescales of approximately 827 and 825 days,
respectively, compared with 1154 and 1099 days under the
SDE-based mDRW model. However, given the relatively large approximate standard errors, these differences are not substantial relative to the estimation uncertainties. The fitted log-likelihoods are 482.66 and 487.86, respectively, with corresponding AIC values of $-915.32$ and $-925.71$, favoring the SDE-based model.}

\textcolor{black}{To investigate the implications of the different dependence
structures, Figure~\ref{fig2} displays the estimated marginal PSDs along the diagonal and squared coherence functions in the lower triangle. Squared coherence between bands $i$ and $j$ is defined as
\[
\gamma_{ij}^{2}(\omega)
=
\frac{|S_{ij}(\omega)|^{2}}
     {S_{ii}(\omega)S_{jj}(\omega)},
\]
and it measures the strength of linear dependence between two
stationary processes at a given temporal frequency
\citep[e.g.,][]{2014A&ARv..22...72U}. Together, these quantities summarize the fitted marginal variability and cross-band dependence in the frequency domain, apart from the mean levels.}

\textcolor{black}{The diagonal panels
show nearly overlapping marginal PSDs for the two models.
This agreement is consistent with their common DRW marginal
specification and similar estimates of the stationary standard
deviations and characteristic timescales. In particular, the
$u$ band exhibits greater variability power than the other
bands, while the remaining bands have relatively similar
marginal PSD amplitudes. Thus, the choice of dependence
structure has relatively little effect on the inferred
band-specific stochastic variability in this example.}

\textcolor{black}{Previous studies have used Gaussian processes to
characterize the temporal variability of individual
AGN light curves, including comparisons of stochastic
covariance kernels \citep{Liu2026} and assessments
of possible periodic variability \citep{Covino2020}.
These studies primarily investigate temporal dependence
within individual light curves, which can also be
characterized through their marginal PSDs.
Our analysis addresses a complementary question:
how dependence among simultaneously observed
photometric bands is represented and inferred.
In particular, similar marginal PSDs under different
multi-output GP formulations do not necessarily imply
similar cross-band dependence.}

\textcolor{black}{The lower-triangular panels of Figure~\ref{fig2}
reveal substantial differences in the inferred cross-band
dependence. 
Under the covariance-based multivariate Mat\'ern model,
coherence generally varies with frequency because the
cross-covariance and marginal covariance functions have
different characteristic temporal structures. For several band pairs, particularly those involving the \(u\) or \(g\) bands and the longer-wavelength bands, the fitted squared coherence is appreciably lower at low frequencies and increases toward unity at higher frequencies. This indicates that a larger fraction of the variability is linearly shared across these bands at shorter timescales (larger frequencies), even though their marginal processes follow DRWs.}

\textcolor{black}{In contrast, the SDE-based mDRW model implies frequency-independent
squared coherence for each band pair. This follows from its
representation of the multiple light curves through band-specific
linear stochastic dynamics driven by correlated Brownian
motions. Although the marginal response timescales differ
across bands, their transfer functions cancel in the squared
coherence, leaving a constant determined by the correlation
between the stochastic drivers. This driving-noise correlation generally differs from the stationary zero-lag correlation between the light curves when their characteristic timescales differ. In this example, the estimated
driving correlations are close to unity, producing nearly
perfect coherence across the displayed frequencies. Thus,
the SDE-based model attributes the observed cross-band
dependence primarily to a strongly shared stochastic driving
process, with band-specific temporal responses.}

\textcolor{black}{
These contrasting results have different astrophysical
implications. The SDE-based mDRW model represents
cross-band dependence through correlated stochastic
drivers and band-specific temporal responses, implying
frequency-independent squared coherence even when
the marginal variability timescales differ. In contrast,
the covariance-based multivariate Mat\'ern model allows
the degree of shared variability to change with temporal
frequency, potentially reflecting different contributions
to cross-band variability across timescales. The present
analysis does not establish which physical mechanism
operates in this quasar, since the inferred coherence
depends on the assumed model and the data provide
limited information about long characteristic timescales.
Nevertheless, the comparison demonstrates that similar
marginal variability estimates can lead to substantially
different inferences about cross-band dependence,
highlighting the importance of dependence-structure
modeling in multi-band AGN studies.
}

\subsection{Reverberation Mapping under the Latent-Process Formulation}
\label{sec42}

Reverberation mapping (RM) provides a powerful means of probing spatially
unresolved structures in active galactic nuclei (AGNs) through their temporal
variability \citep{Blandford1982,Peterson1993,PetersonHorne2004}. In emission-line
reverberation mapping, variations in the ionizing continuum are followed by
corresponding variations in broad emission lines after a characteristic time
delay associated with the light-travel time across the broad-line region (BLR). Unlike the multi-band photometric variability
considered in Section~\ref{sec41}, the dependence between the
continuum and emission-line light curves is modeled through a
physically motivated convolution operator rather than specified
directly by a cross-covariance function. This example therefore
provides a natural astronomical application of the latent-process
framework introduced in Section~\ref{sec22}.

\begin{figure}
\begin{center}
\includegraphics[scale = 0.26]{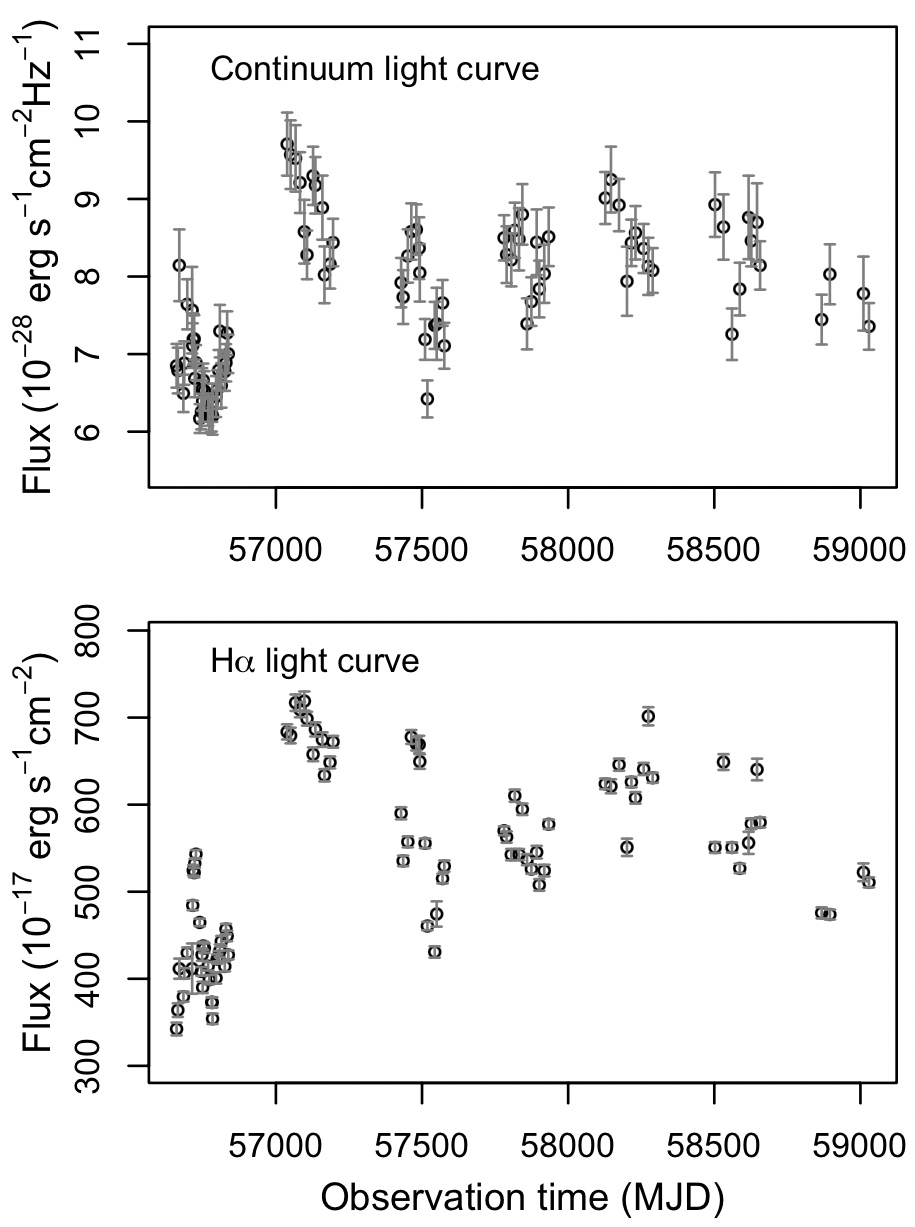}
\caption{
Observed continuum (upper panel) and H$\alpha$ emission-line
(lower panel) light curves of the SDSS-RM target RM840
\citep{shen2024}. The reported $1\sigma$ measurement
uncertainties are displayed as gray error bars, although they
are too small to be readily visible.
}
\label{fig4}
\end{center}
\end{figure}

We illustrate this framework using one target, RM840, from the SDSS
Reverberation Mapping (SDSS-RM) project
\citep{shen2024}. The data consist of the SDSS spectroscopic $r$-band
continuum light curve together with the H$\alpha$ emission-line
light curve. Figure~\ref{fig4} displays the two observed light curves. The
median cadence of both the continuum and emission-line
observations is 11.97 days, while the median $1\sigma$
measurement uncertainties are
$0.32\times10^{-28}\,\mathrm{erg\,s^{-1}\,cm^{-2}\,Hz^{-1}}$
for the continuum and
$6.44\times10^{-17}\,\mathrm{erg\,s^{-1}\,cm^{-2}}$
for the H$\alpha$ emission-line light curve.

Following the latent-process framework introduced in
Section~\ref{sec22}, we assume that the latent continuum
variability is described by a zero-mean DRW process,
\[
Z(t)
\sim
\mathrm{DRW}(0,\sigma,\tau),
\]
where $\sigma$ and $\tau$ denote the diffusion coefficient
(short-timescale variability amplitude) and characteristic
timescale, respectively. The corresponding stochastic differential equation is
$$
dZ(t) = -\frac{1}{\tau}Z(t)+\sigma dW(t),
$$
where \(W(t)\) is a standard Brownian motion.

The latent continuum light curve is
modeled as
\[
X_c(t)
=
\mu_c
+
Z(t),
\]
where $\mu_c$ denotes the mean continuum flux. Throughout this
section, the subscripts $c$ and $\ell$ denote the continuum and
emission-line processes, respectively.

The latent H$\alpha$ emission-line light curve is generated by
applying a linear operator to the latent continuum process,
\[
X_\ell(t)
=
\mu_\ell
+
\alpha_\ell
\int
\Psi(u)
Z(t-u)\,du,
\]
where $\mu_\ell$ denotes the mean emission-line flux,
$\alpha_\ell$ is the response amplitude, and
$\Psi(\cdot)$ denotes the transfer function that accounts for  how the time lags are distributed across the broad line region.

To investigate the
effect of the linear operator on the resulting covariance structure, we consider two commonly adopted transfer functions.
The first is the top-hat transfer function used in JAVELIN
\citep{zu2011an,zu2016app},
\[
\Psi_{\rm TH}(u)
=
\frac{1}{w_{\rm TH}}
I
\left(
\tau_{0,{\rm TH}}
-
\frac{w_{\rm TH}}{2}
\le
u
\le
\tau_{0,{\rm TH}}
+
\frac{w_{\rm TH}}{2}
\right),
\]
where $\tau_{0,{\rm TH}}$ denotes the mean reverberation lag and
$w_{\rm TH}$ denotes the width of the response. The second is the
Gaussian transfer function 
\citep{2014A&ARv..22...72U, shen2015, li2016},
\[
\Psi_{\rm G}(u)
=
\frac{1}{\sqrt{2\pi}\sigma_{\rm G}}
\exp
\left[
-
\frac{
(u-\tau_{0,{\rm G}})^2
}
{
2\sigma_{\rm G}^2
}
\right],
\]
where $\tau_{0,{\rm G}}$ denotes the mean reverberation lag and
$\sigma_{\rm G}$ denotes the standard deviation.

\textcolor{black}{Our objective is not to benchmark lag estimates against
existing reverberation mapping methods, which may differ
in their additional modeling assumptions and estimation
procedures. Instead, by holding the latent continuum
model and likelihood-based estimation procedure fixed
within a common multi-output GP framework, we isolate
the effect of the transfer-function shape on the inferred
dependence structure and its astronomical interpretation.}

The corresponding PSDs follow immediately from the 
result in Eq.~\eqref{eq:psd_latent}. The transfer function
determines the frequency-domain representation of the observed
emission-line variability through the squared modulus of its
Fourier transform. The frequency-domain properties of top-hat
and Gaussian transfer functions have been discussed previously
in the context of X-ray reverberation mapping
\citep{2014A&ARv..22...72U}. Here, we place these
transfer-function models within the latent-process multi-output
GP framework, derive the covariance structures they induce for a
latent DRW process, and compare their likelihood-based fits.

Given a single latent DRW process, the resulting cross-covariance functions depend on the choice of transfer function. Accordingly, the marginal PSD of the observed
emission-line process is derived as
\[
S_\ell(\omega)
=
\alpha_\ell^2
\left|
\widehat{\Psi}(\omega)
\right|^2
S_Z(\omega),
\]
where $S_Z(\omega)$ denotes the Lorentzian PSD of the latent DRW
process defined in Eq.~\eqref{eq:psd_latentdrw}. The top-hat and Gaussian transfer functions yield two different marginal PSDs for the same emission-line light curve:
\[
S_{\ell}^{\rm TH}(\omega)
=
\alpha_\ell^2
\operatorname{sinc}^2
\left(
\frac{\omega w_{\rm TH}}{2}
\right)
S_Z(\omega),
\]
and
\[
S_{\ell}^{\rm G}(\omega)
=
\alpha_\ell^2
\exp
\left(
-\omega^2\sigma_{\rm G}^2
\right)
S_Z(\omega),
\]
respectively, where
\[
\operatorname{sinc}(x)
=
\frac{\sin(x)}{x},
\qquad
\operatorname{sinc}(0)=1.
\]
A detailed
derivation is provided in Appendix~\ref{appendixc}. Thus, given the same latent stochastic continuum variability,
the transfer function determines how that variability is
redistributed across temporal frequencies in the emission-line light curve.

With the latent process and transfer-function assumptions,
the likelihood function  is fully
determined by the cross-covariance functions in
Eq.~\eqref{eq:latent_crosscov}. In a preliminary analysis,
the response-width estimates approach the lower boundary
of 0.05 days under both transfer-function models, suggesting
that the temporal spread of the response is weakly
constrained by these data. To focus on the effect of the
transfer-function shape, we therefore fix the Gaussian
response standard deviation at $\sigma_{\rm G}=5$ days
and the top-hat width at
$w_{\rm TH}=\sqrt{12}\sigma_{\rm G}\approx17.3$ days.
These choices give the two transfer functions the same
delay variance, with widths comparable to the typical
within-season observational cadence. Thus, the fixed
widths provide a controlled comparison of the two
transfer-function shapes rather than independently
estimated response widths. We then estimate all
remaining model parameters by maximum likelihood.
The resulting parameter estimates and their
Hessian-based standard errors are reported in
Table~\ref{tab:rmfit}.

\begin{table}[t]
\centering
\small
\setlength{\tabcolsep}{6pt}
\renewcommand{\arraystretch}{1}
\caption{Maximum likelihood estimates and Hessian-based standard
errors for the two latent-process models. The models share the
same latent DRW formulation and differ only in the choice of
transfer function. The Akaike information criterion (AIC)
and mean CPU time per likelihood evaluation (in milliseconds), measured under the same
computational conditions, are also reported for comparison.}
\label{tab:rmfit}

\begin{tabular}{l cc c cc}
\hline
& \multicolumn{2}{c}{Top-hat} && \multicolumn{2}{c}{Gaussian} \\
\cline{2-3} \cline{5-6}
Parameter & Estimate & SE && Estimate & SE \\
\hline
$\mu_c$           & 8.06   & 0.17  && 8.05   & 0.17  \\
$\sigma$          & 0.16   & 0.02  && 0.16   & 0.02  \\
$\tau$ (days)     & 51.43  & 13.98 && 51.13  & 13.84 \\
$\mu_\ell$        & 534.57 & 21.50 && 536.37 & 21.33 \\
$\alpha_\ell$     & 128.75 & 12.87 && 129.76 & 13.06 \\
$\tau_0$ (days)   & 136.51 & 0.11  && 138.68 & 0.19  \\[0.3em]
\hline
AIC & \multicolumn{2}{c}{1112.13} && \multicolumn{2}{c}{1100.98} \\
CPU time & \multicolumn{2}{c}{0.26} && \multicolumn{2}{c}{0.26} \\
\hline
\end{tabular}
\end{table}

The estimated latent DRW parameters, $\sigma$ and $\tau$, are
very similar under the two transfer-function models, indicating
that the inferred continuum variability is largely insensitive
to the assumed operator. Likewise, the estimated continuum and
emission-line means ($\mu_c, \mu_\ell$), and response amplitudes ($\alpha_\ell$) are broadly comparable under the two models. The estimated mean reverberation lags \(\tau_0\) are 136.51 and 138.68 days under the top-hat and Gaussian models, respectively, differing by only 2.17 days. Although the within-model standard errors are small, the difference between the estimates illustrates the sensitivity of lag inference to the assumed transfer-function shape.

\begin{figure}
\begin{center}
\includegraphics[scale = 0.24]{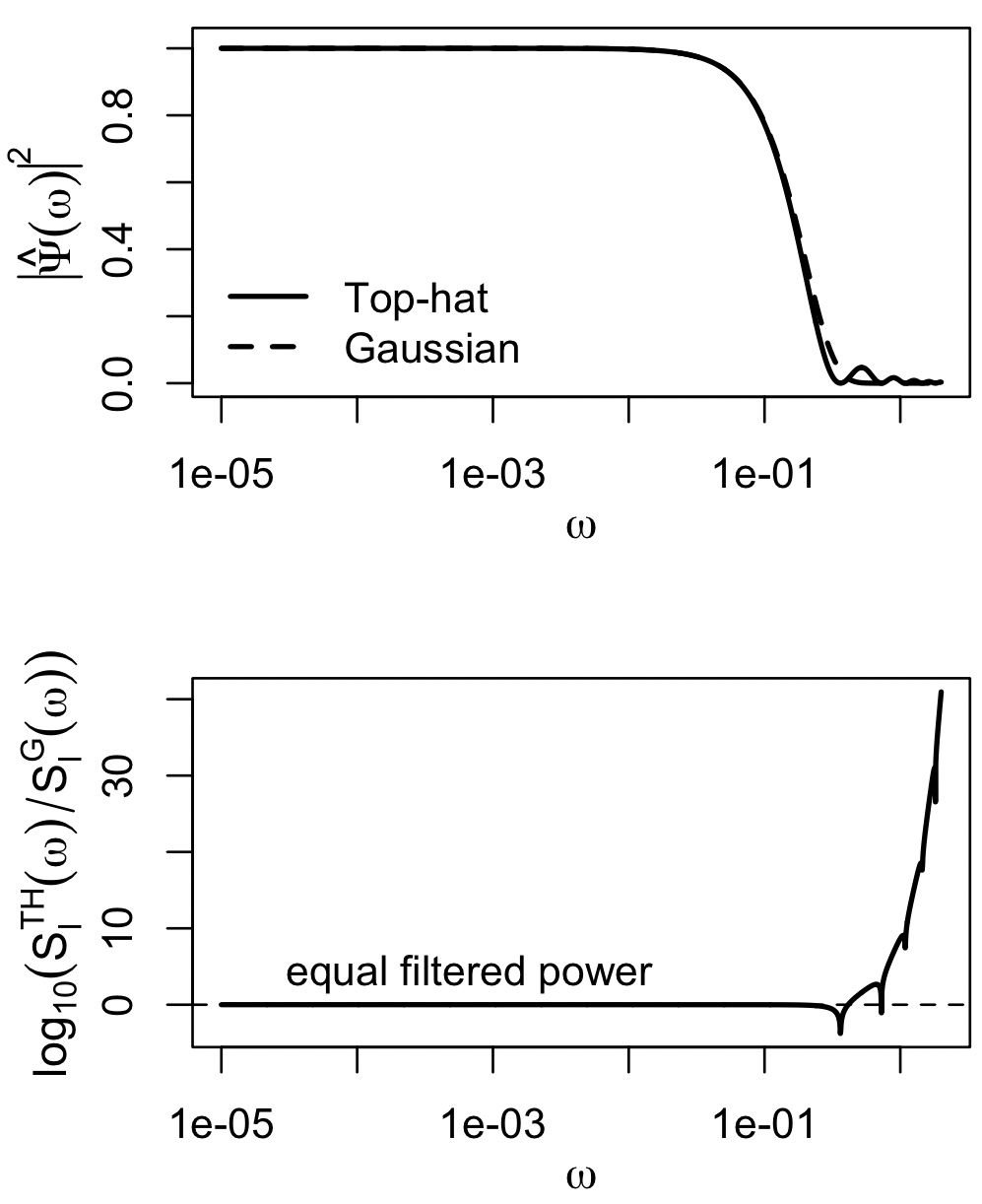}
\caption{
Frequency-domain comparison of the fixed-width top-hat and
Gaussian transfer-function models. The upper panel displays the
squared magnitudes of the corresponding Fourier transforms. The
lower panel shows the base-10 logarithm of the ratio of the fitted
H$\alpha$ emission-line PSDs. The horizontal dashed line denotes
equal filtered power under the two models. Values above zero indicate that the top-hat model predicts
greater emission-line power at the corresponding temporal
frequency, whereas values below zero favor the Gaussian model.
}
\label{fig7}
\end{center}
\end{figure}

Given these estimates, we discuss how the different transfer functions affect the dependence structure and the resulting stochastic variability of the emission-line light curve in the frequency domain.
Figure~\ref{fig7} compares the frequency responses induced by the fitted
fixed-width top-hat and Gaussian transfer-function models. The
upper panel displays the squared magnitudes of the corresponding
Fourier transforms, i.e., $|
\widehat{\Psi}_i(\omega)|^2$   in Eq.~\eqref{eq:psd_latent}, which determine how the latent continuum
variability is filtered before being observed in the emission
line. At low temporal frequencies, both responses remain close to
unity, indicating that long-timescale continuum variability is
transmitted similarly under the two models. Their behavior
diverges at higher frequencies: the Gaussian response decreases
smoothly and monotonically, whereas the top-hat response exhibits
the oscillatory sidelobes characteristic of the sinc function
associated with a finite response window. Consequently, since the two
models yield nearly identical estimates of the latent DRW
parameters,  the
stochastic behavior of the latent emission-line variability is
governed by the assumed temporal response of the broad-line
region through the transfer function.

The lower panel of Figure~\ref{fig7} displays the logarithm of
the ratio of the fitted emission-line PSDs,
\[
\log_{10}
\left\{
\frac{
S_{\ell}^{\rm TH}(\omega)
}{
S_{\ell}^{\rm G}(\omega)
}
\right\}.
\]
Values close to zero indicate that the two models predict nearly
identical emission-line variability at the corresponding
temporal frequency. The ratio remains close to zero throughout
the low-frequency range. At higher frequencies, the Gaussian response decays smoothly and exponentially, whereas the top-hat response exhibits oscillatory sidelobes with algebraically decaying amplitude. Consequently, the two models can yield substantially different predictions for unresolved short-timescale emission-line variability.

From an astrophysical perspective, the transfer function
describes how continuum variations are redistributed in time
before being reprocessed into emission-line radiation by the
broad-line region. A top-hat transfer function assumes a finite
range of delays with approximately equal contributions, whereas
a Gaussian transfer function represents a smoothly distributed
delay field. Although these assumptions lead to nearly identical
fits to the observed light curves, they imply different
frequency responses and hence different stochastic descriptions
of the unresolved emission-line variability. In the present
data, these differences arise primarily at large temporal frequencies
that are only weakly constrained by the observations,
suggesting that the mean reverberation lag is well identified
whereas the detailed shape of the transfer function remains
comparatively difficult to infer.

From a statistical perspective, the Gaussian transfer-function
model is favored over the top-hat model according to the
Akaike information criterion \citep[AIC;][]{akaike1974new},
\textcolor{black}{with nearly identical average likelihood evaluation times (0.26 milliseconds)}.
The AIC difference of 11.15 indicates substantial support
for the Gaussian model relative to the top-hat model among
the two fixed-width models considered here. However, this
preference is conditional on the specified response widths
and should not be interpreted as definitive evidence for
a Gaussian physical transfer function, since the true
response of the broad-line region may differ from both
models. Model selection is therefore most informative
when physically motivated transfer functions are compared
within a common statistical framework.

\textcolor{black}{The top-hat and Gaussian transfer functions both imply
unit squared coherence between the latent continuum and
emission-line processes,
\[
\gamma_{c\ell}^2(\omega)
=
\frac{|S_{c\ell}(\omega)|^2}
{S_{cc}(\omega)S_{\ell\ell}(\omega)}
=1,
\]
at frequencies where the denominator is positive; see
Appendix~\ref{appendixc3} for a derivation. By definition,
unit squared coherence indicates a perfect linear
relationship between the two processes at each frequency.
This result holds for any stationary latent continuum
process passed through a deterministic linear convolution
operator, provided that no additional independent
stochastic variability is introduced into the emission-line
process. Thus, although the two transfer functions induce
different emission-line PSDs and continuum-line
cross-covariance structures, both imply perfect squared
coherence because the continuum is the sole source of
stochastic variability in the emission-line process.}

More generally, this example illustrates that dependence
structures describe not only the observed covariance of the data
but also the unresolved stochastic variability between
observations. Consequently, different dependence structures may
lead to different scientific interpretations even when they
provide nearly identical fits to the observed light curves.
These differences are expected to become increasingly important
as the observational cadence becomes sparse relative to the
characteristic timescale of the unresolved variability, because
the assumed dependence structure plays a progressively larger
role in determining the latent trajectories, predictive
uncertainties, and ultimately the inferred reverberation lag.
Quantifying these effects under realistic survey cadences
provides an important direction for future work.
\\

Together, the two examples illustrate that dependence
structures play different scientific roles depending on
the underlying research question. In the SDSS Stripe~82
example, covariance-based and SDE-based formulations
provide different descriptions of cross-band dependence,
even when their marginal PSDs are nearly identical.
In the reverberation mapping example, different
transfer functions induce distinct marginal PSDs and
cross-covariance structures, despite implying the same
unit squared coherence between the latent continuum
and emission-line processes.

These results emphasize that the dependence structure
is not merely a mathematical specification of
cross-covariances but an essential modeling decision
that influences the scientific interpretation of
multi-output GPs. The appropriate formulation should
therefore be guided by the scientific mechanism and
inferential objectives, much as the choice of a
covariance kernel is central to univariate GP modeling.

\section{Concluding Remarks}\label{sec5}

This paper has examined three complementary approaches to
modeling dependence in multi-output Gaussian processes:
covariance-based, SDE-based, and latent-process formulations.
The two astronomical examples illustrate that different
dependence structures can lead to different scientific
interpretations even when they provide similar descriptions
of the observed variability. In the SDSS Stripe~82 example,
the covariance-based multivariate Mat\'ern and SDE-based
multivariate DRW models yield nearly identical marginal
PSDs but substantially different squared coherence
functions. In the reverberation mapping example, the
top-hat and Gaussian transfer functions imply different
emission-line PSDs and cross-covariance structures,
despite both producing unit squared coherence, reflecting
the assumption that the continuum is the sole source of
stochastic variability in the emission-line process.
These results emphasize that the choice of dependence
structure is a scientific modeling decision rather than
merely a choice of covariance parameterization.

These comparisons also provide guidance for choosing
dependence structures according to scientific objectives.
When the goal is to characterize cross-band covariance
or frequency-domain dependence, covariance-based approaches
allow direct specification of the relevant statistical
relationships. When the goal is to describe joint stochastic
evolution across bands, SDE-based approaches provide a
natural formulation through coupled dynamics or correlated
driving processes. When dependence arises from a hypothesized
physical relationship, such as the delayed response of
emission-line variability to continuum fluctuations,
latent-process approaches allow that relationship to be
incorporated explicitly through suitable transformations.
These considerations do not establish a general ranking
among the approaches, nor do they imply that the approaches
are mutually exclusive. Rather, they illustrate how the
scientific questions of interest can guide the construction
and interpretation of multi-output GP models.

\subsection{Beyond the DRW Process}

Although the illustrations in this work focus on the
DRW, its single characteristic
timescale may be insufficient to describe more complex
 variability. In particular, higher-order
temporal dependence, multiple characteristic timescales,
and more complicated spectral behavior may require
richer stochastic processes.

The three dependence formulations considered here are
not restricted to the DRW. Covariance-based models can
incorporate more flexible temporal kernels, including
Mat\'ern kernels, mixtures of kernels, and quasi-periodic
covariance functions. Latent-process formulations can
similarly accommodate richer underlying stochastic
processes while retaining physically motivated linear
operators that describe relationships among the outputs.
In the SDE-based formulation, a natural extension is
provided by multivariate continuous-time autoregressive
moving-average  processes
\citep{marquardt2007mCARMA,schlemm2012multivariate,
BROCKWELL2013217} for analyzing multi-band light curves
of AGNs \citep{schmidlkofer2026mcarma,
schmidlkofer2026mcarma2}. These processes generalize the
multivariate DRW by allowing higher-order stochastic
dynamics, multiple characteristic timescales, and
more flexible marginal and cross-spectral structures.

Extending the three formulations beyond the DRW would
allow temporal variability and cross-output dependence
to be modeled with greater flexibility. Such extensions
are particularly relevant when the scientific objectives
require characterizing departures from DRW behavior or
identifying dependence structures that cannot be
adequately represented by a single characteristic
timescale.

\subsection{Computational Challenges}\label{sec52}

Regardless of the chosen dependence formulation, richer
stochastic models introduce additional computational
challenges. If $N$ denotes the total number of
observations across all outputs, direct likelihood
evaluation for a dense Gaussian process generally
requires $\mathcal{O}(N^3)$ computation and
$\mathcal{O}(N^2)$ memory, making exact inference
impractical for large astronomical surveys.

Several complementary strategies offer promising
directions for scalable inference. State-space
representations permit exact Kalman filtering and
smoothing for models with finite-dimensional Markov
representations, including DRW and CARMA processes
\citep{kelly2009,kelly2014,sarkka2013,
tak2017bayesian,2020AJ....160..265H,meyer2023}.
Conditional approximations such as the Vecchia
approximation \citep{vecchia1988,katzfuss2021}
and sparse variational Gaussian process methods
\citep{titsias2009,hensman2013,hensman2015,
alvarez2010,alvarez2011} can substantially reduce
computational cost while retaining much of the
flexibility of GP models. An alternative direction
is simulation-based inference (SBI), which learns
approximations to posterior distributions from
simulated data
\citep{papamakarios2019,cranmer2020,
papamakarios2021,lueckmann2021}. SBI is particularly
attractive when likelihood evaluation is analytically
or computationally intractable.

\subsection{Uncertainty Quantification}\label{sec53}

The examples considered in this paper primarily use
maximum likelihood estimation to illustrate the
scientific interpretation of different dependence
structures. Although approximate Hessian-based
standard errors are reported for the estimated model
parameters, uncertainty in derived quantities,
including covariance functions, PSDs, coherence
functions, and transfer functions, has not been
fully propagated throughout the analysis.

One possible extension is to propagate the asymptotic
covariance matrix of the maximum likelihood estimator
to derived quantities using the multivariate delta
method. This approach would provide approximate
confidence intervals or confidence bands for PSDs,
coherence functions, and other covariance or
frequency-domain summaries. Alternatively, one could
draw parameter vectors from a multivariate normal
approximation,
$N(\hat{\theta},\hat{\Sigma})$,
where $\hat{\theta}$ denotes the maximum likelihood
estimator and $\hat{\Sigma}$ its estimated covariance
matrix, and evaluate the corresponding derived
quantities for each draw. Such approximations require
care when the likelihood is poorly conditioned,
parameters are near the boundaries of their admissible
regions, or the estimated covariance matrix is not
positive definite. Parametric bootstrap procedures
provide another option for assessing uncertainty
when asymptotic approximations are unreliable.

A Bayesian approach provides an alternative treatment
of uncertainty by assigning prior distributions to
the model parameters and characterizing their joint
posterior distribution. Posterior samples can be
propagated through the covariance and spectral
representations to obtain credible bands for PSDs,
coherence functions, transfer functions, and other
derived quantities. This approach naturally accounts
for parameter dependence and is particularly useful
when comparing scientific interpretations under
different dependence structures. Developing efficient
Bayesian inference for multi-output GP models,
particularly in conjunction with the computational
strategies discussed in Section~\ref{sec52},
remains an important direction for future work.
\\

In conclusion, the scientific interpretation of
multi-output astronomical time series depends not
only on the temporal variability assumed for each
output but also on how dependence among outputs
is constructed. Covariance-based, SDE-based, and
latent-process formulations offer complementary
ways to represent this dependence, with their
relative usefulness depending on the scientific
objectives and assumptions of the analysis.
Extending these approaches to richer stochastic
processes, together with scalable inference and
uncertainty quantification, will provide greater
flexibility for analyzing multi-band variability
and reverberation mapping data from modern
astronomical surveys.

\section*{Acknowledgements}

S.~Das, L.~Shi, Y.~Homayouni, and H.~Tak were supported by the
Rising Researcher Program 2025--26 (ICDS-RR25-027363 and
ICDS-RR25-027377) through Penn State's Institute for
Computational and Data Sciences (RRID:SCR-025154) and gratefully
acknowledge the institute for providing access to its
computational research infrastructure (RRID:SCR-026424).
S.~Das, L.~Shi, and H.~Tak were also  supported by the
Penn State Center for Astrostatistics and Astroinformatics.

H.~Tak acknowledges partial support from the Brain Pool
Fellowship Program 2026--27 of the National Research Foundation
of Korea (RS-2026-25551883). J.-H.~Woo acknowledges support 
from the Basic Science Research Program through the National
Research Foundation of Korea  (RS-2026-25487266).

\software{R (R Core Team, 2026, version 4.6.1), and a Python package, mcarma \citep[][version 1.2.0]{mcarma2026}. The R and Python scripts used in this work are publicly available at the
corresponding author's GitHub repository:
\url{https://github.com/hyungsuktak/mgp}.
}

\appendix


\section{Derivation of the Cross-Spectral Densities}
\label{appendixa}

For a weakly stationary multi-output GP, define $K_{ij}(u)
=
\operatorname{Cov}\{X_i(t+u),X_j(t)\},$
where $u$ denotes the time lag. The corresponding cross-spectral density is
\begin{equation}\label{eq:app_fourier}
S_{ij}(\omega)
=
\int_{-\infty}^{\infty}
K_{ij}(u)e^{-i\omega u}\,du.
\end{equation}
We derive the cross-spectral densities for the three dependence constructions considered in Section~\ref{sec3}. The marginal PSDs follow by setting $i=j$.

\subsection{Covariance-Based Multivariate Mat\'ern Process}

For the multivariate Mat\'ern construction with $\nu_{ij}=1/2$, the cross-covariance function in Eq.~\eqref{eq:matern_drw_cross} is
\[
K_{ij}(u)
=
\rho_{ij}s_i s_j
\exp\left(-\frac{|u|}{\tau_{ij}}\right).
\]
Since this function is even in $u$, its Fourier transform becomes
$$
S_{ij}(\omega)
=
2\rho_{ij}s_i s_j
\int_0^\infty
e^{-u/\tau_{ij}}\cos(\omega u)\,du=
\frac{2\rho_{ij}s_i s_j\tau_{ij}}
{1+\omega^2\tau_{ij}^2},
$$
where we use
\[
\int_0^\infty e^{-au}\cos(bu)\,du
=
\frac{a}{a^2+b^2},
\qquad a>0.
\]
This recovers Eq.~\eqref{eq:psd_matern}.

\subsection{SDE-Based Multivariate DRW Process}

For the multivariate DRW, the stationary cross-covariance function in Eq.~\eqref{eq:mdrw_cross} is
\[
K_{ij}(u)
=
\rho_{ij}s_i s_j
\begin{cases}
e^{-u/\tau_i}, & u\geq 0,\\
e^{u/\tau_j}, & u<0.
\end{cases}
\]
Splitting the Fourier integral at zero gives
$$
S_{ij}(\omega)
=
\rho_{ij}s_i s_j
\left[
\int_0^\infty
e^{-u/\tau_i}e^{-i\omega u}\,du
+
\int_{-\infty}^0
e^{u/\tau_j}e^{-i\omega u}\,du
\right]
=
\rho_{ij}s_i s_j
\left[
\frac{\tau_i}{1+i\omega\tau_i}
+
\frac{\tau_j}{1-i\omega\tau_j}
\right]
=
\frac{\rho_{ij}s_i s_j(\tau_i+\tau_j)}
{(1+i\omega\tau_i)(1-i\omega\tau_j)}.
$$
The resulting cross-spectrum is generally complex-valued when $\tau_i\neq\tau_j$. 

\subsection{Convolution-Based Latent-Process Model}

Consider the reverberation-mapping construction with a single zero-mean stationary latent process $Z(t)$,
\[
X_i(t)
=
\int_{-\infty}^{\infty}
\Psi_i(a)Z(t-a)\,da,
\]
where $\Psi_i$ is a real-valued transfer function. Let
\[
K^{(Z)}(u)
=
\operatorname{Cov}\{Z(t+u),Z(t)\},
\qquad
S_Z(\omega)
=
\int_{-\infty}^{\infty}
K^{(Z)}(u)e^{-i\omega u}\,du.
\]
The cross-covariance between the latent band-specific light curves is
\[
K_{ij}(u)
=
\int_{-\infty}^{\infty}
\int_{-\infty}^{\infty}
\Psi_i(a)\Psi_j(b)
K^{(Z)}(u-a+b)\,da\,db.
\]
Taking its Fourier transform and interchanging the integrals yields
\begin{align*}
S_{ij}(\omega)
&=
\int\!\!\int
\Psi_i(a)\Psi_j(b)
\left[
\int
K^{(Z)}(u-a+b)e^{-i\omega u}\,du
\right]da\,db.
\end{align*}
With the change of variable $v=u-a+b$, the inner integral becomes
$e^{-i\omega a}e^{i\omega b}S_Z(\omega).$
Defining the Fourier transform of each transfer function as
\[
\widehat{\Psi}_i(\omega)
=
\int_{-\infty}^{\infty}
\Psi_i(a)e^{-i\omega a}\,da,
\]
we obtain
\[
S_{ij}(\omega)
=
\widehat{\Psi}_i(\omega)
\overline{\widehat{\Psi}_j(\omega)}
S_Z(\omega),
\]
which recovers Eq.~\eqref{eq:psd_latent}. In particular,
\[
S_{ii}(\omega)
=
|\widehat{\Psi}_i(\omega)|^2S_Z(\omega).
\]

\section{Parameter Estimates for the SDSS Multi-band AGN Analysis}
\label{appendixb}

Table~\ref{tab:sdss_estimates} reports the maximum likelihood
estimates and approximate standard errors for the covariance-based
multivariate Mat\'ern model with $\nu_{ij}=1/2$ and the SDE-based
multivariate DRW model fitted to the SDSS Stripe~82 quasar
(dbID: 1540). The stationary standard deviations $s_j$ and
characteristic timescales $\tau_j$ determine the marginal PSDs,
while the zero-lag correlations $\rho_{ij}$ characterize the
dependence between latent light curves at the same time.
For the SDE-based model, $\rho_{ij}$ is calculated from the
estimated correlations $r_{ij}$ of the driving Brownian motions as
$\rho_{ij}=2r_{ij}\sqrt{\tau_i\tau_j}/(\tau_i+\tau_j)$.

\begin{table}[b!]
\begin{center}
\caption{
Maximum likelihood estimates for the covariance-based multivariate
Mat\'ern model and SDE-based multivariate DRW model fitted to the
SDSS Stripe~82 quasar (dbID: 1540). Approximate standard errors,
obtained using inverse-Hessian approximations and the delta method,
are reported in parentheses. Both Hessian calculations exhibit
numerical instability, so the standard errors should be interpreted
with caution.
}
\label{tab:sdss_estimates}
\small
\setlength{\tabcolsep}{4pt}

\begin{tabular}{lcccccccc}
\hline\hline
Model & $s_u$ & $s_g$ & $s_r$ & $s_i$ & $s_z$
      & $\tau_u$ & $\tau_g$ & $\tau_r$ \\
\hline
Mat\'ern
& 0.166 (0.043) & 0.060 (0.013) & 0.065 (0.014)
& 0.049 (0.011) & 0.049 (0.012)
& 425 (199) & 353 (149) & 689 (377) \\
mDRW
& 0.159 (0.034) & 0.056 (0.010) & 0.067 (0.012)
& 0.055 (0.012) & 0.054 (0.013)
& 441 (163) & 351 (112) & 698 (233) \\
\hline
\end{tabular}

\vspace{1em}

\begin{tabular}{lcccccccc}
\hline\hline
Model & $\tau_i$ & $\tau_z$ & $\rho_{ug}$ & $\rho_{ur}$
      & $\rho_{ui}$ & $\rho_{uz}$ & $\rho_{gr}$ & $\rho_{gi}$ \\
\hline
Mat\'ern
& 827 (415) & 825 (457)
& 0.992 (0.024) & 0.943 (0.083)
& 0.873 (0.107) & 0.870 (0.129)
& 0.899 (0.087) & 0.825 (0.105) \\
mDRW
& 1154 (466) & 1098 (513)
& 0.993 (0.014) & 0.974 (0.028)
& 0.895 (0.062) & 0.904 (0.077)
& 0.944 (0.026) & 0.846 (0.055) \\
\hline
\end{tabular}

\vspace{1em}

\begin{tabular}{lcccc}
\hline\hline
Model & $\rho_{gz}$ & $\rho_{ri}$
      & $\rho_{rz}$ & $\rho_{iz}$ \\
\hline
Mat\'ern
& 0.822 (0.130) & 0.974 (0.032)
& 0.971 (0.050) & 1.000 (0.003) \\
mDRW
& 0.857 (0.077) & 0.969 (0.022)
& 0.975 (0.034) & 1.000 (0.004) \\
\hline
\end{tabular}

\end{center}
\end{table}

\section{Derivation of the Emission-Line Power Spectral Densities}
\label{appendixc}

Appendix~\ref{appendixb} shows that, under the latent-process
framework, the PSD matrix is determined by the Fourier
transforms of the linear operators acting on the latent
stochastic process. For the continuum reverberation mapping model
considered in Section~4, the emission-line process is
\[
X_\ell(t)
=
\mu_\ell
+
\alpha_\ell
\int
\Psi(u)
Z(t-u)\,du,
\]
where $\mu_\ell$ denotes the mean emission-line flux,
$\alpha_\ell$ is the response amplitude, and
$\Psi(\cdot)$ is the transfer function. Because the PSD depends
only on the covariance (or equivalently, the centered process),
the constant mean $\mu_\ell$ has no effect on the spectrum.
Since multiplication by the constant $\alpha_\ell$ scales the
Fourier transform by the same factor, the corresponding
emission-line PSD becomes
\[
S_\ell(\omega)
=
\alpha_\ell^2
\left|
\widehat{\Psi}(\omega)
\right|^2
S_Z(\omega).
\]
Therefore, it remains only to derive the Fourier transforms of
the transfer functions considered in
Section~\ref{sec42}.

For completeness, we summarize the standard
Fourier transforms of the top-hat and Gaussian transfer
functions \citep[e.g.,][]{2014A&ARv..22...72U} and then
specialize the resulting frequency-domain filters to the latent
DRW process considered in this paper.

\subsection{Top-Hat Transfer Function}

The normalized top-hat transfer function is
\[
\Psi_{\rm TH}(u)
=
\frac{1}{w_{\rm TH}}
I
\left(
\tau_{0,\rm TH}-\frac{w_{\rm TH}}2
\le u
\le
\tau_{0,\rm TH}+\frac{w_{\rm TH}}2
\right).
\]

Its Fourier transform is
$$
\widehat{\Psi}_{\rm TH}(\omega)=
\frac1{w_{\rm TH}}
\int_{\tau_0-w/2}^{\tau_0+w/2}
e^{-i\omega u}\,du=
e^{-i\omega\tau_{0,\rm TH}}
\frac{2\sin(\omega w_{\rm TH}/2)}
{\omega w_{\rm TH}}
=
e^{-i\omega\tau_{0,\rm TH}}
\operatorname{sinc}
\left(
\frac{\omega w_{\rm TH}}2
\right),
$$
where $\operatorname{sinc}(x)=\sin x / x$.

Since the phase factor has unit modulus,
\[
\left|
\widehat{\Psi}_{\rm TH}(\omega)
\right|^2
=
\operatorname{sinc}^2
\left(
\frac{\omega w_{\rm TH}}2
\right),
\]
and therefore
\[
S_\ell^{\rm TH}(\omega)
=
\alpha_\ell^2
\operatorname{sinc}^2
\left(
\frac{\omega w_{\rm TH}}2
\right)
S_Z(\omega).
\]

\subsection{Gaussian Transfer Function}

The normalized Gaussian transfer function is
\[
\Psi_{\rm G}(u)
=
\frac1{\sqrt{2\pi}\sigma_{\rm G}}
\exp
\left[
-
\frac{
(u-\tau_{0,\rm G})^2
}
{
2\sigma_{\rm G}^2
}
\right].
\]

Its Fourier transform is the Fourier transform of a shifted
Gaussian density,
\[
\widehat{\Psi}_{\rm G}(\omega)
=
e^{-i\omega\tau_{0,\rm G}}
\exp
\left(
-\frac{\omega^2\sigma_{\rm G}^2}{2}
\right).
\]

Consequently,
\[
\left|
\widehat{\Psi}_{\rm G}(\omega)
\right|^2
=
\exp
\left(
-\omega^2\sigma_{\rm G}^2
\right),
\]
giving
\[
S_\ell^{\rm G}(\omega)
=
\alpha_\ell^2
\exp
\left(
-\omega^2\sigma_{\rm G}^2
\right)
S_Z(\omega).
\]

\paragraph{Role of the Mean Reverberation Lag.}

The mean reverberation lag enters the Fourier transform through
the phase factor
$e^{-i\omega\tau_0},$
which has unit modulus. Consequently, the lag does not affect
the marginal emission-line PSD. Instead, it appears in the phase
of the continuum--emission-line cross-spectrum and therefore
governs the frequency-dependent time delay between the continuum
and emission-line variability.

\subsection{Squared Coherence under the Latent-Process RM Model}
\label{appendixc3}

We derive the squared coherence between the latent
continuum and emission-line processes under the
linear convolution model introduced in
Section~\ref{sec42}. Since the mean levels do not
affect the spectral densities, we consider the
zero-mean stochastic components,
\[
X_c(t)-\mu_c=Z(t),
\qquad
X_\ell(t)-\mu_\ell
=
\alpha_\ell\int\Psi(u)Z(t-u)\,du.
\]

Let $S_Z(\omega)$ denote the PSD of the latent
continuum process and $\widehat{\Psi}(\omega)$
the Fourier transform of the transfer function.
The marginal PSDs are
\[
S_{cc}(\omega)=S_Z(\omega),
\qquad
S_{\ell\ell}(\omega)
=
\alpha_\ell^2
|\widehat{\Psi}(\omega)|^2S_Z(\omega).
\]
The corresponding cross-spectral density is
\[
S_{c\ell}(\omega)
=
\alpha_\ell
\overline{\widehat{\Psi}(\omega)}
S_Z(\omega),
\]
where the overline denotes complex conjugation,
under the cross-spectral convention used here.

Consequently, the squared coherence is
$$
\gamma_{c\ell}^2(\omega)
=
\frac{|S_{c\ell}(\omega)|^2}
{S_{cc}(\omega)S_{\ell\ell}(\omega)}
=
\frac{
\alpha_\ell^2
|\widehat{\Psi}(\omega)|^2
S_Z^2(\omega)
}{
S_Z(\omega)
\alpha_\ell^2
|\widehat{\Psi}(\omega)|^2
S_Z(\omega)
}
=1.
$$
This equality holds at frequencies where the
denominator is positive. In particular, squared
coherence is undefined at the spectral zeros of
the top-hat transfer function.

Thus, both the top-hat and Gaussian transfer
functions imply unit squared coherence between
the latent continuum and emission-line processes.
This reflects the assumption that the continuum
is the sole source of stochastic variability in
the emission-line process, with no additional
independent stochastic component. Although the
transfer functions produce different marginal
PSDs and cross-covariance structures, the
continuum and emission-line processes remain
perfectly linearly related at each frequency
where coherence is defined.

\bibliography{reference}
\bibliographystyle{aasjournalv7}



\end{document}